# Full control of electric and magnetic light-matter interactions through a plasmonic nanomirror on a near-field tip

*Benoît Reynier[1\*], Eric Charron[1], Obren Markovic[1], Xingyu Yang[1], Bruno Gallas[1], Alban Ferrier[2], Sébastien Bidault[3] and Mathieu Mivelle[1\*]*

[1] Sorbonne Université, CNRS, Institut des NanoSciences de Paris, INSP, 75005 Paris, France
[2] Chimie ParisTech, PSL University, CNRS, Institut de Recherche de Chimie Paris, 75005 Paris, France
[3] Institut Langevin, ESPCI Paris, Université PSL, CNRS, 75005 Paris, France
*mathieu.mivelle@sorbonne-universite.fr

**Light-matter interactions are often considered governed by the electric optical field only, leaving aside the magnetic component of light. However, the magnetic part plays a determining role in many optical processes from light and chiral-matter interactions, photon-avalanching to forbidden photochemistry, making the manipulation of magnetic processes extremely relevant. Here, by creating a standing wave using a plasmonic nanomirror we manipulate the spatial distributions of the electric and magnetic fields and their associated local density of states, allowing the selective control of the excitation and emission of electric and magnetic dipolar transitions. This control allows us to image, in 3D, the electric and magnetic nodes and anti-nodes of the fields' interference pattern. It also enables us to enhance specifically photoluminescence from quantum emitters excited only by the magnetic field, and to manipulate their quantum environment by acting on the excitation fields solely, demonstrating full control of magnetic and electric light-matter interactions.**

Manipulating light-matter interactions at the nanoscale has revolutionized many scientific fields. Whether it be in biology, with ever more sensitive diagnostics platforms,[1, 2] medicine with targeted therapies,[3, 4] chemistry with higher efficiency catalysis,[5, 6] or physical optics with ever more exotic manipulations of these interactions.[7-11] Nevertheless, most of the systems developed to date have aimed at manipulating the electric component of light, leaving aside its magnetic counterpart. Indeed, light-matter interactions are often considered driven by the electric optical field alone, ignoring the magnetic component of light. However, this magnetic component plays a key role in many optical processes, such as chiral light-matter interactions,[12] ultrasensitive detection,[13] enhancement of Raman optical activity,[14] photon-avalanching,[15] or forbidden photochemistry,[16] which makes the manipulation of magnetic processes extremely important. In fact, lately, the manipulation of 'magnetic light'-matter interactions was also achieved to a certain extent. For instance, luminescence mediated by magnetic transition dipoles was controlled and enhanced by manipulating the magnetic local density of states (LDOS) through metallic layers acting as mirrors[17-22] or with resonant dielectric[23-32] and plasmonic[33-36] nanostructures. It was also demonstrated that a Bessel beam could selectively excite a magnetic dipole transition through the magnetic field of light.[37]

Here, we introduce a new platform made of a plasmonic nanomirror creating a standing wave pattern to manipulate the spatial distributions of the electric and magnetic fields and the associated local densities of optical states. With this platform, we demonstrate the selective excitation of electric (ED) or magnetic (MD) dipolar transitions and selectively collect the luminescence emitted by ED or MD transitions. This control allows us to image, in 3D, the electric and magnetic nodes and anti-nodes of the fields' interference pattern. It also allows us to specifically enhance the luminescence of the quantum emitter by magnetic excitation only and to manipulate the quantum environment of the latter by acting on the excitation fields only, thus demonstrating total control of the magnetic and electrical light-matter interactions.

**Results**

For this purpose, a plasmonic nano-antenna is fabricated at the tip of an aluminum-coated tapered optical fiber in a Scanning Near-Field Optical Microscope (SNOM) and acts as a nanomirror when excited from the far-field to create a standing wave (figure 1). This electromagnetic field is used to excite a $Eu^{3+}$ doped $Y_2O_3$ nanoparticle, whose position can be scanned at the nanoscale in 3D under the SNOM tip, allowing a dynamical control of the interactions.

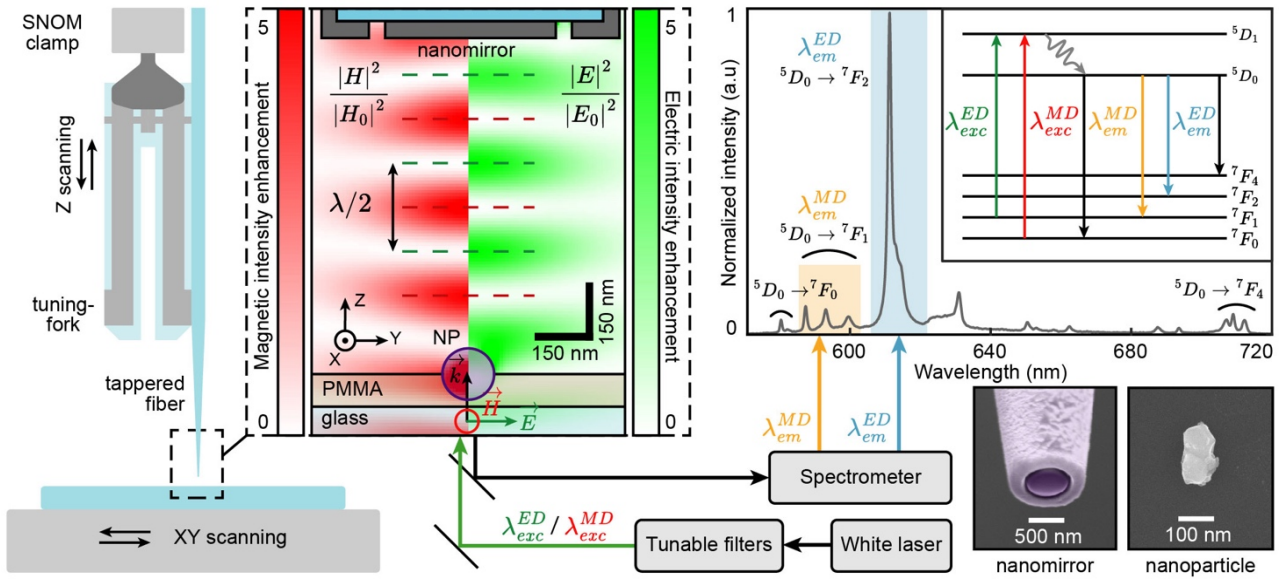

**Figure 1**: Principal of the experiment. A plasmonic nanomirror fabricated at the tip of a tapered fiber and placed on a scanning near-field optical microscope is brought near a $Y_2O_3$ nanoparticle doped with $Eu^{3+}$ ions fixed on a glass substrate by a PMMA polymer layer. The nanopositioning capabilities of the tip and the piezoelectric stage on which the particle is placed allow us to perfectly control the 3D position of the $Eu^{3+}$ ions under the plasmonic nanomirror. The excitation of the coupled nanomirror-particle system is done from the sample via a microscope objective (x100, 1.3 NA) and by a tunable laser in wavelength. The same objective collects the emitted photons, and after filtering from the laser light, a spectrometer provides the luminescence spectrum for each position of the particle under the nanomirror. As a zoom in below the nanomirror are displayed numerical simulations of the standing wave generated by the plasmonic nanomirror when it is excited by an incident wave, linearly polarized along Y, launched from the substrate side distant by 1000 nm. The interferences of the magnetic intensity of the standing wave at $\lambda_{exc}^{MD}$ are on the left side, in red, and on the right side, in green, those of the electric intensity at $\lambda_{exc}^{ED}$. Both intensities are normalized by the amplitude of the incident field. The dotted lines are guides for the eye showing the spatial separation of the electric and magnetic anti-nodes in the standing wave. The purple circle indicates the $Eu^{3+}$-doped particle partially embedded in a thin PMMA layer, which is spin-coated on a glass substrate. In the inset is represented the emission spectrum of $Eu^{3+}$ ions in the $Y_2O_3$ matrix with the magnetic and electric transitions of interest highlighted in yellow and blue, respectively, and the partial band diagram of $Eu^{3+}$ ions. On the latter are shown the electric ($\lambda_{exc}^{ED}$) and magnetic ($\lambda_{exc}^{MD}$) transitions at the excitation and respectively at the emission ($\lambda_{em}^{ED}$, $\lambda_{em}^{MD}$).

$Eu^{3+}$ ions are known to exhibit pure electric and magnetic transitions in the visible spectrum, both in terms of excitation[37] and emission[18] (partial band diagram in the inset of figure 1). The excitation of the ED (at $\lambda_{exc}^{ED}$ = 532 nm) and MD (at $\lambda_{exc}^{MD}$ = 527.5 nm) transitions is then performed by a white laser coupled to a series of tunable filters, allowing the reduction of the laser spectrum to a bandwidth of only 2 nm. The luminescence of the ED (at $\lambda_{em}^{ED}$ = 610 nm) and MD (at $\lambda_{em}^{MD}$ = 590 nm) transitions of the Eu$^{3+}$ ions is then collected by the same objective, filtered from the laser light, and measured by a spectrometer. The emission spectrum of europium ions is shown in figure 1. By tuning the position of the nanoparticle within the standing wave, we can thus selectively excite it with the E or H field and selectively collect the signal emitted by the ED and MD transitions. Therefore, we have access to the 3D distributions of the electromagnetic fields and of the local densities of optical states that act on the quantum emitters.

Figure 1 shows the theoretical spatial distributions (see methods) of the electric and magnetic fields generated by the standing wave beneath the plasmonic nano-mirror at the $\lambda_{exc}^{ED}$ and $\lambda_{exc}^{MD}$ wavelengths, respectively. We observe that the electric and magnetic nodes and anti-nodes do not overlap spatially. A maximum E field corresponds to a minimum H field and vice versa. Also, inside the anti-nodes, the field intensities are increased by a factor of five compared to the incident wave. Finally, due to the different continuity conditions at the interfaces, we can see that the two components of light do not penetrate the doped nanoparticle in the same way, with a clear predominance of the magnetic field inside the latter. Interestingly, this means that the H field will excite the MD transitions more efficiently than their ED counterparts by the E field, but also that the E and H excitations take place at different positions within the nanoparticle.

The luminescence intensity *L* of the europium-doped nanoparticle is proportional to the excitation intensity according to the following equation: $L = \sigma |A|^2 \eta Q$, where $\sigma$ is the absorption cross-section, *A* is the electric or magnetic excitation field, $\eta$ is the collection efficiency, and *Q* the quantum yield. Figure 2a provides the luminescence collected at $\lambda_{em}^{ED}$ when exciting the particle at $\lambda_{exc}^{ED}$ and $\lambda_{exc}^{MD}$ for different antenna-particle distances *Z* and normalized with respect to the luminescence intensities without the nanomirror. We observe that the signals do not overlap spatially: the maxima and minima for these two excitations are almost inversed, in excellent agreement with the theoretical results expected from the excitation of the particle by the E or H field of light. These measurements thus indicate that the evolution of *L* as a function of *Z* follows directly the evolution of the excitation probability and that *Q* and $\eta$ have a negligible influence on the spatial distributions of the luminescence intensities. Importantly, since *Q* and $\eta$ are independent of the nature of the excitation process (MD or ED) and only depend on *Z*, it is possible to divide the luminescence enhancement measured at $\lambda_{exc}^{ED}$ by the luminescence enhancement measured at $\lambda_{exc}^{MD}$ and recover directly the ratio between the intensity enhancements of the E and H fields, providing a quantitative agreement between measurements and theory (see Supporting Information and figure S2).

As a control experiment, the measurement is also performed using a 200 nm diameter nanoparticle filled with fluorescent molecules (figure 2b, see methods and SI). In this case, magnetic transitions are negligible compared to their electrical counterpart, and the absorption spectrum overlaps with both the $\lambda_{exc}^{ED}$ and $\lambda_{exc}^{MD}$ wavelengths (see figure S1). For fluorescent nanospheres, the curves are perfectly superimposed, and the signal follows a purely electric excitation. This measurement confirms that the luminescence collected in figure 2a for a $\lambda_{exc}^{ED}$ excitation represents the spatial distribution of the E field intensity in the standing wave and that the signal for a $\lambda_{exc}^{MD}$ excitation maps the magnetic field. Furthermore, we observe that the fluorescence intensity is enhanced by a factor of 3 and 2.5 for, respectively, the excitation by the E and H fields compared to the signal collected without the antenna. This measurement provides the first demonstration of an enhanced luminescence signal from quantum emitters excited specifically by the magnetic component of light.

Moreover, using the SNOM nano-positioning capabilities, the luminescence of $Eu^{3+}$ ions can be collected in a volume under the nanomirror, providing a 3D spatial reconstruction of the E and H field intensities of the standing wave as shown in Figure 2c. Here, the E and H nodes and anti-nodes are observed as lobes of the standing wave because of the nanoscale size of the plasmonic mirror. This is the first 3D image providing, in parallel, the intensities of the electric and magnetic components of light.

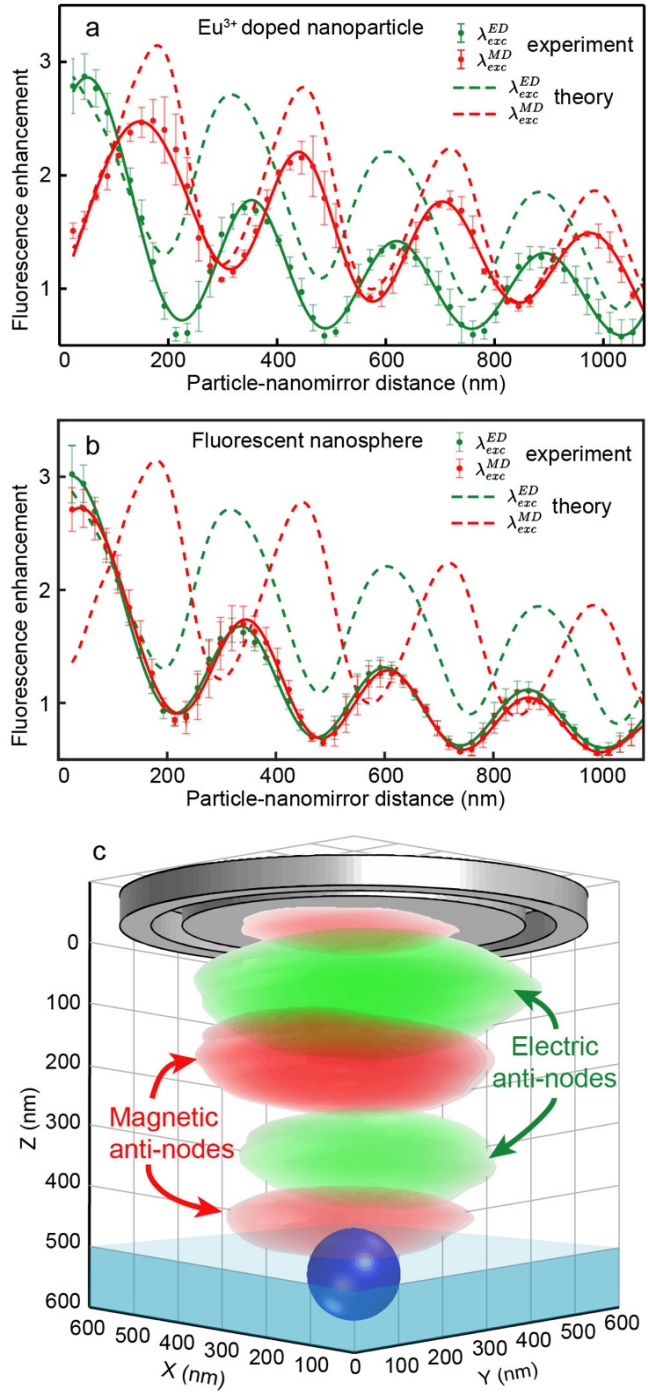

***Figure 2.*** *Optical characterization of the standing wave. a) Increase of the luminescence intensities emitted by the Eu³⁺-doped particle and collected by the spectrometer for excitation wavelengths at $\lambda_{exc}^{ED}$ (in green) and $\lambda_{exc}^{MD}$ (in red) and for different Z positions of the particle under the nanomirror. b) Increase in fluorescence intensity emitted from nanospheres filled with fluorescent molecules (see methods) for different Z positions under the nanomirror and excited at $\lambda_{exc}^{ED}$ (in green) and $\lambda_{exc}^{MD}$ (in red). In a) and b), the points correspond to the average values of the experimental data normalized by the signal without the antenna, the solid curves are polynomial fits serving as guides for the eye, and the dashed curves correspond to numerical calculations of the expected signal for an excitation by the magnetic field, in red, or the electric field, in green, of light. The error bars*

*correspond to the standard deviation. c) 3D image of the electric, in green, and magnetic, in red, nodes and anti-nodes of the electromagnetic standing wave generated under the nanomirror.*

Finally, by tuning the excitation wavelength and studying separately the ED and MD emission intensities, we study how the plasmonic nano-mirror modifies the spontaneous emission rates for an electric or magnetic excitation. Since the emitted photons originate from the same excited state, we can infer the $\beta^{ED}$ and $\beta^{MD}$ branching ratios by considering any other transitions and non-radiative decay channels as losses:[20]

$$\beta^{ED} = \frac{L^{ED}}{L^{ED} + L^{MD}} = 1 - \beta^{MD} \quad (1)$$

It is then possible to determine the relative local densities of states experienced by the ED (at $\lambda_{em}^{ED}$) and MD (at $\lambda_{em}^{MD}$) transitions as:[20]

$$\tilde{\rho}^{ED} = \frac{\rho^{ED}}{\rho^{ED} + \rho^{MD}} = \frac{\beta_{NM}^{ED}/\beta_0^{ED}}{\beta_{NM}^{ED}/\beta_0^{ED} + \beta_{NM}^{MD}/\beta_0^{MD}} = 1 - \tilde{\rho}^{MD} \quad (2)$$

with $\beta_{NM}$ and $\beta_0$ representing the branching ratios with and without nanomirror, respectively. Figure 3 provides the radiative electric LDOS when exciting the particle using the E and H field, respectively, for different nanomirror-particle distances. Interestingly, these two LDOSes, although measured at the same positions and thus in the same photonic environment, do not overlap spatially. The explanation can be found in the non-finite size of the Eu³⁺ doped nanoparticle. Indeed, depending on the component of light that interacts with the particle, the position of the excited ions will not spatially overlap because of a different spatial distribution of the fields within the particle as shown in Figure 1. The emitting ions will therefore be at different positions corresponding to a different LDOS. Thus, by changing the nature of the exciting field, it is possible to turn on or off some ions and probe different quantum environments. These subtle variations are in good agreement with the theory when the LDOS, inferred from the photoluminescence measurements, is balanced by the distribution of the excitation fields in the particle (figures 3).

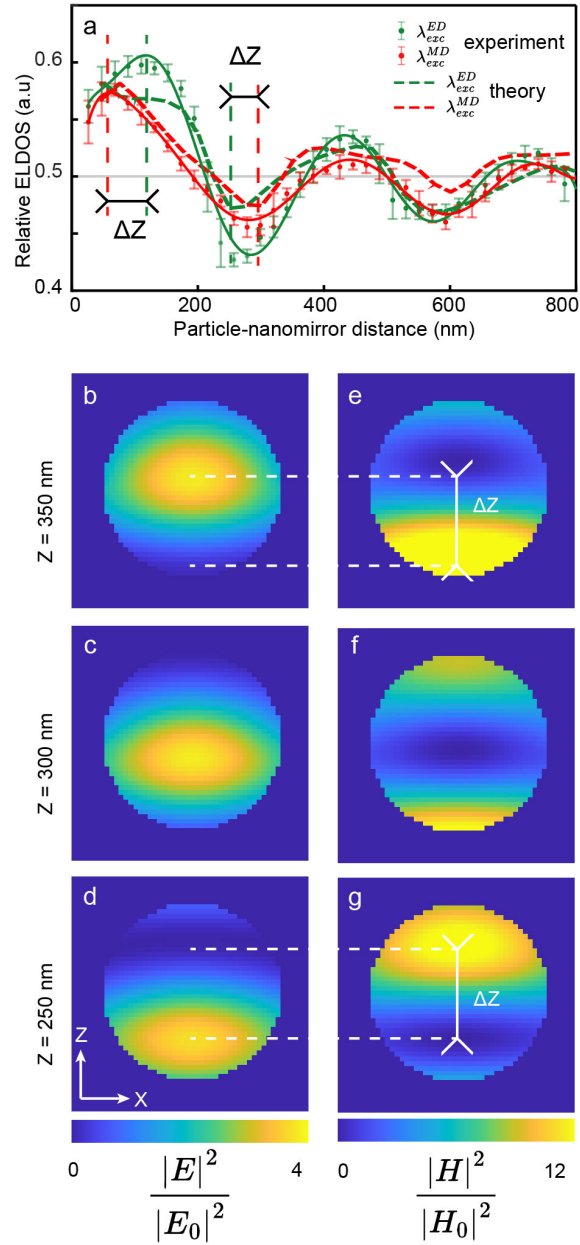

**Figure 3**. LDOS change through field excitation. a) Relative electric LDOS as a function of the particle-nanomirror distance when the $Eu^{3+}$ ions are excited with the magnetic field, in red, or the electric field, in green. The points correspond to the average values of the experimental data, the solid curves are polynomial fits serving as guides for the eye, and the dotted curves correspond to the numerical calculations. The error bars correspond to the standard deviation; Z=0 is chosen as the top part of the doped particle. Theoretical distribution of (b-d) electric and (e-g) magnetic optical fields inside the nanoparticle, normalized by the incident wave and for different Z positions of the particle under the nanomirror (indicated on the left side). The diameter of the nanoparticle is 150 nm.

**Conclusion**

In conclusion, through a new platform, we demonstrated that by generating a standing wave with a nanomirror at the end of a SNOM tip, we could perfectly control the electric and magnetic interactions of light with quantum emitters, both in terms of the excitation probability and of the spontaneous decay channels. This manipulation allowed us to provide the first experimental 3D image of the electric and magnetic nodes and anti-nodes of a standing wave. Furthermore, we demonstrated an increase in the emission of a quantum emitter after specific excitation of its magnetic transition dipole, and we showed how, by this full control of the interactions, we could manipulate the spontaneous emission of an emitter only by acting on its excitation. This research opens the way to many photonic applications involving a contribution from the optical magnetic field, such as chiral light-matter interactions, photochemistry, manipulation of magnetic processes, and new schemes in quantum computing or nonlinear processes, among others.

**Methods**

Nanomirror fabrication:

The fabrication of the nanostructured tips is done in several steps. First, a P-2000 puller from Sutter pulls an optical fiber to manufacture the fibered tip. A layer of 120 nm aluminum is then deposited on the perimeter of the fibered tip. A focused ion beam then cuts the tip to a core diameter of about 800 nm. A second 50 nm thick aluminum layer is deposited by thermal evaporation on the processed end part of the tip. Finally, a circular groove is generated by FIB to produce a nanomirror of 550 nm in diameter.

Nanoparticles synthesis:

2% Eu:$Y_2O_3$ nanoparticle of 200 ± 50 nm average diameter were prepared by homogeneous precipitation.[38] In this method, an aqueous nitrate solution of $Y(NO_3)_3$. $6H_2O$ (99.9% pure, Alfa Aesar), $Eu(NO_3)_3$. $6H_2O$ (99.99% pure, Reacton) was mixed with an aqueous urea solution $(CO(NH_2)_2, > 99\%$ pure, Sigma) in a Teflon reactor. The pH inside the reactor was then slowly increased during a 24 h thermal treatment at 85°C by the urea decomposition. The metal and urea concentrations were 7.5 mmol L$^{-1}$ and 3 mol L$^{-1}$ respectively. After cooling, a white precipitate of amorphous yttrium hydroxycarbonate ($Eu^{3+}$: $Y(OH)CO_3$.n H2O) was collected by centrifugation and washed at least 3 times with water and absolute ethanol. That amorphous powder was then converted to highly crystalline Eu:$Y_2O_3$ nanoparticles by 24 h calcination treatment at 1000°C (rate of 3 °C min$^{-1}$). The body-centered cubic $Y_2O_3$ structure (Ia-3 space group) of the particles was confirmed by X-ray diffraction with no evidence for other parasitic phases.

Simulations:

The simulations were done by the finite difference time domain (FDTD) software Lumerical. The aluminum nanomirror placed at the end of a fibered tip has a diameter of 550 nm for a metal thickness of 50 nm. The latter is separated from the rest of the tip by a gap of 50 nm. The optical index of the glass fiber tip is 1.46, and the permittivity of aluminum used in the simulation was experimentally measured by ellipsometry. Under the tip is placed a $Y_2O_3$ nanoparticle of 150 nm diameter and optical index of 1.94. The latter sits on a glass substrate and is half embedded in a 100 nm layer of PMMA with index 1.5 and 1.46, respectively. The total size of the simulation window is 2x2x3 um$^3$ and the finest mesh defining the most detailed parts is 4 nm.

Two types of simulations are performed, one with a source made of a Gaussian beam coming from the substrate at wavelengths $\lambda_{exc}^{ED}$ and $\lambda_{exc}^{MD}$, the other with electric and magnetic dipoles placed inside the nanoparticle and emitting at wavelengths $\lambda_{em}^{ED}$ and $\lambda_{em}^{MD}$.

The use of the Gaussian beam allows describing the excitation of the Eu$^{3+}$ doped nanoparticle by the standing wave. On the other hand, the collection of the powers emitted by the dipoles and radiated in the far field allows for determining the local electric and magnetic density of optical states experienced by the ions.[20]

Setup:

Excitation of Eu$^{3+}$-doped nanoparticles is performed by a white laser (NKT Photonics K90-110-10), filtered by a combination of interference filters (Semrock BrightLine FF01-532/18-25 and Spectrolight FWS-B-F06), in order to reduce the spectral bandwidth to 2 nm while maintaining high laser power. First, the excitation light is sent through the inverted microscope (Olympus IX73) to an oil objective (Olympus PLN 100x Oil Immersion, NA 1.30) that focuses the light under the nanomirror. The same optical path then collects the luminescence, and a dichroic mirror (Semrock BrightLine FF552-Di01-25x36) sends it to a spectrometer (Sol Instruments MS5204i). The luminescence spectra are then measured with a CCD camera (Andor iDus 401 CCD). The optical near-field microscope (NT-MDT-Integra) is placed on the inverted microscope, and the tip supporting the nanomirror is glued on a tuning fork vibrating at a frequency of 32kHz. The approach and the feedback loop of the tip in the near field are made by monitoring the phase of the oscillation of the tuning fork. Next, the tip is aligned on the center of the immersion lens, and the particle is scanned under it thanks to a piezo stage (Piezoconcept), allowing a nanometric displacement.

Sample preparation:

Eu$^{3+}$ doped nanoparticle samples. After cleaning by sonic bath and plasma cleaner, a 110 nm layer of PMMA is deposited by spin-coating (3% weight - 4000 rpm) on glass coverslips then annealed at 180° for one minute to evaporate the excess solvent and homogenize the layer. To make the layer hydrophilic, the sample is once again treated with plasma cleaner, reducing the PMMA layer to 80 nm thickness. The doped nanoparticles are then deposited by spin-coating on the sample, and new annealing is performed at 180° for one minute to allow the nanoparticles to embed in the polymer.

Fluorosphere samples. After a sonic bath and plasma cleaner, a 3% diluted PVA solution containing fluorescent beads (F8763 fluorescent beads, Thermofisher) is spin-coated at 1000 rpm on a glass coverslip.

**Data availability**

The data and datasets that support the plots within this paper and other findings of this study are available from the corresponding author upon reasonable request.

**Author contributions**

M.M supervised the study. B.R, E.C and O.M performed the experiments. B.R and X.Y performed the numerical study. A.F synthesised the $Eu^{3+}$-doped nanoparticles. B.R, B.G, S.B and M.M analysed the data. All the authors discussed the results and contributed to writing the manuscript.

**Competing interests**

The authors declare no competing interests.

**Additional information**

Supplementary information

# Supplementary Information

# Full control of electric and magnetic light-matter interactions through a plasmonic nanomirror on a near-field tip

*Benoît Reynier[1*], Eric Charron[1], Obren Markovic[1], Xingyu Yang[1], Bruno Gallas[1], Alban Ferrier[2], Sébastien Bidault[3] and Mathieu Mivelle[1*]*

[1] Sorbonne Université, CNRS, Institut des NanoSciences de Paris, INSP, 75005 Paris, France

[2] Chimie ParisTech, PSL University, CNRS, Institut de Recherche de Chimie Paris, 75005 Paris, France

[3] Institut Langevin, ESPCI Paris, Université PSL, CNRS, 75005 Paris, France

*mathieu.mivelle@sorbonne-universite.fr

## 1. $Eu^{3+}$-doped $Y_2O_3$ nanoparticles and fluorescent spheres

The europium ions $Eu^{3+}$ can be excited through several transitions. In particular, the $^7F_0 \rightarrow {}^5D_1$ transition at $\lambda_{exc}^{MD} = 527$ nm and the $^7F_1 \rightarrow {}^5D_1$ transition at $\lambda_{exc}^{ED} = 532$ nm have been shown to be mediated by magnetic and electric transition dipoles, respectively. Figure S1a represents an excitation spectrum around these two transitions.

As a control, we also used 200 nm diameter nanospheres filled with fluorescent molecules (F8763 fluorescent beads, Thermofisher) for which the electric transition dipole dominates its magnetic counterpart. The emission and excitation spectra of these nanospheres are shown in figure S1b. We can see that these nanospheres can be excited at both $\lambda_{exc}^{MD}$ and $\lambda_{exc}^{ED}$.

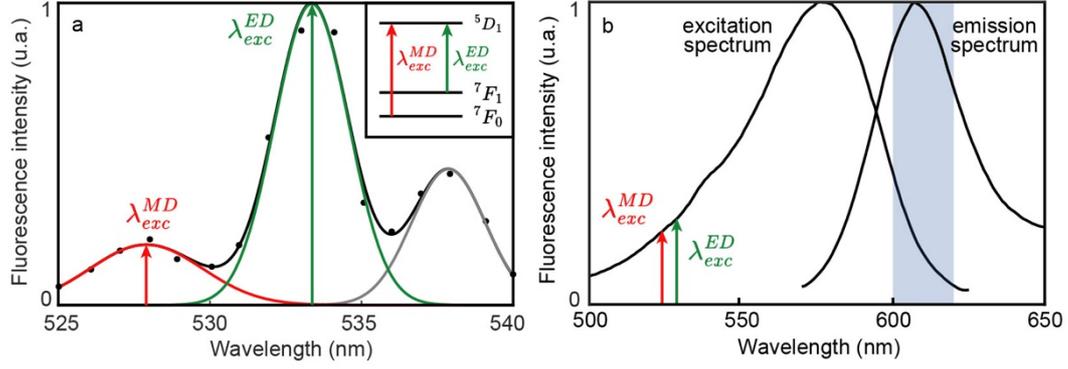

**Figure S1.** a) Excitation spectrum of Eu$^{3+}$-doped nanoparticles. The excitation wavelengths $\lambda_{exc}^{MD}$ and $\lambda_{exc}^{ED}$ are shown in red and green, respectively. In the inset is represented the partial band diagram of these two transitions b) Excitation and emission spectra of the fluorescent nanobeads. The two excitation wavelengths $\lambda_{exc}^{MD}$ and $\lambda_{exc}^{ED}$ are shown in red and green, respectively. The shaded part corresponds to the collected fluorescence.

Experimental methods for optical measurements:

The excitation spectrum (Figure S1a) is measured by scanning the excitation wavelength in steps of 1nm and collecting the luminescence signal from the transitions $^5D_0 \rightarrow {}^7F_2$ at $\lambda_{em}^{ED}$.

## 2. Theoretical background: Excitation study

The luminescence L at the emission wavelength $\lambda_i$ for the i transition (i=ED or MD), excited by the field A (with A the electric E or magnetic H optical field) can be defined as:

$$L(A, \lambda_i) = \sigma(A) \times |A|^2 \times \eta(\lambda_i) \times Q(\lambda_i) \quad (S1)$$

where, $\sigma(A)$ is the absorption cross-section, $|A|^2$ is the electric or the magnetic field intensity, $\eta(\lambda_i)$ and $Q(\lambda_i)$ are the collection efficiency and the quantum yield of the transition, respectively.

To demonstrate that the oscillations of the electric and magnetic optical intensities follow exactly the measured oscillations in the collected luminescence signal represented in figure 2, figure S2 represents the ratio $R_{E/H}$ of the theoretical and experimental signals:

$$R_{E/H} = \frac{\frac{L_{NM}(E, \lambda_i)}{L_0(E, \lambda_i)}}{\frac{L_{NM}(H, \lambda_i)}{L_0(H, \lambda_i)}} \quad (S2)$$

Where $L_{NM}(A, \lambda_i)$ and $L_0(A, \lambda_i)$ are the luminescence signals collected for an excitation by a field A at $\lambda_i$, with or without the nano-mirror, respectively.

Therefore $R_{E/H}$ can be written as:

$$\frac{L_{NM}(E,\lambda_i)}{L_0(E,\lambda_i)} \times \frac{L_0(H,\lambda_i)}{L_{NM}(H,\lambda_i)} = \frac{\sigma(E)}{\sigma(E)} \frac{|E_{NM}|^2}{|E_0|^2} \frac{\eta_{NM}(\lambda_i)}{\eta_0(\lambda_i)} \frac{Q_{NM}(\lambda_i)}{Q_0(\lambda_i)} \times \frac{\sigma(H)}{\sigma(H)} \frac{|H_0|^2}{|H_{NM}|^2} \frac{\eta_0(\lambda_i)}{\eta_{NM}(\lambda_i)} \frac{Q_0(\lambda_i)}{Q_{NM}(\lambda_i)} = \frac{|E_{NM}|^2}{|E_0|^2} \times \frac{|H_0|^2}{|H_{NM}|^2} \quad (S3)$$

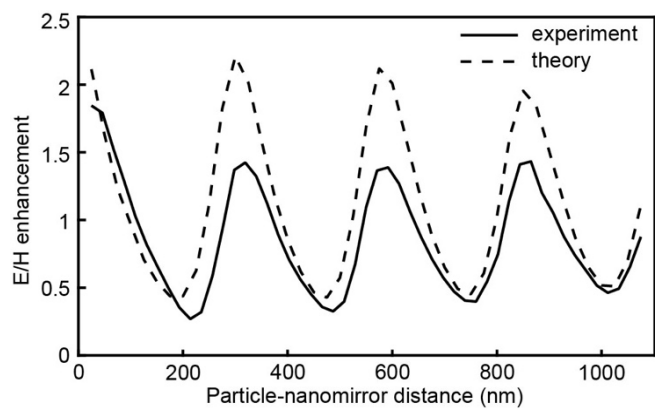

**Figure S2**. *Theoretical and experimental ratios of the luminescence signals for electric and magnetic excitation and for different particle-mirror distances.*

We can see in figure S2 that there is an excellent theoretical and experimental agreement between the ratios of the luminescence signals for electric and magnetic excitations.